\documentclass[showkeys,twocolumn,nofootinbib]{revtex4-1}
\usepackage{subscript}
\usepackage{natbib}
\usepackage{amssymb}
\usepackage{color}
\usepackage{amsfonts}
\usepackage{textcomp}
\newcommand{\ped}[1]{\ensuremath{_{\rm #1}}}
\newcommand{\apex}[1]{\ensuremath{^{\rm #1}}}
\definecolor{blue}{rgb}{0,0,0}
\usepackage{float}
\usepackage[caption = false]{subfig}
\usepackage{graphicx}
\definecolor{link}{RGB}{57,106,177}
\usepackage[hypertex]{hyperref}
\hypersetup{
    colorlinks=true,
    linkcolor=link,
    citecolor=link,
    urlcolor=link
}

\begin{document}

\title{Mapping multi-valley Lifshitz transitions induced by field-effect doping in strained MoS$_2$ nanolayers}

\author{Erik Piatti}
\author{Davide Romanin}
\author{Renato S. Gonnelli}
\email{renato.gonnelli@polito.it}
\affiliation{Department of Applied Science and Technology, Politecnico di Torino, 10129 Torino, Italy}

\begin{abstract}
Gate-induced superconductivity at the surface of nanolayers of semiconducting transition metal dichalcogenides (TMDs) has attracted a lot of attention in recent years, thanks to the sizeable transition temperature, robustness against in-plane magnetic fields beyond the Pauli limit, and hints to a non-conventional nature of the pairing. A key information necessary to unveil its microscopic origin is the geometry of the Fermi surface hosting the Cooper pairs as a function of field-effect doping, which is dictated by the filling of the inequivalent valleys at the K/K\apex{\prime} and Q/Q\apex{\prime} points of the Brillouin Zone. Here, we achieve this by combining Density Functional Theory calculations of the bandstructure with transport measurements on ion-gated $2H$-MoS\ped{2} nanolayers. We show that, when the number of layers and the amount of strain are set to their experimental values, the Fermi level crosses the bottom of the high-energy valleys at Q/Q\apex{\prime} at doping levels where characteristic kinks in the transconductance are experimentally detected. We also develop a simple 2D model which is able to quantitatively describe the broadening of the kinks observed upon increasing temperature. We demonstrate that this combined approach can be employed to map the dependence of the Fermi surface of TMD nanolayers on field-effect doping, detect Lifshitz transitions, and provide a method to determine the amount of strain and spin-orbit splitting between sub-bands from electric transport measurements in real devices.
\end{abstract}


\maketitle

\section{Introduction}\label{sec:introduction}

The control of the electronic ground state of nanolayers of layered materials in a field-effect transistor (FET) has been a significant breakthrough in recent years, with the ionic gating technique allowing the quasi-continuous exploration of different electronic phases at reduced dimensionality \cite{YeNatMater2010, YeScience2012, JoNanoLett2015, ShiSciRep2015, YuNatNano2015, SaitoACSNano2015, LiNature2016, WangNature2016, XiPRL2016, OvchinnikovNatCommun2016, LeiPRL2016, ZengNanoLett2018, DengNature2018, WangNatNano2018}. Transition metal dichalcogenides (TMDs) have been among the most studied materials, thanks to their complex phase diagrams which often host both superconductivity (SC) and charge density waves (CDW) \cite{KlemmBook2012, KlemmPhysC2015}. Among these, archetypal layered semiconductor molybdenum disulphide ($2H$-MoS\ped{2}, see Fig.\ref{figure:material_structure}a) has been extensively investigated after the discovery that a gate-tunable SC state, with a sizeable critical temperature $T_c$ up to $\sim 11$ K, could be induced either through electrostatic ion accumulation at the surface \cite{YeScience2012, BiscarasNatCommun2015} or electrochemical ion intercalation in the bulk \cite{ZhangNanoLett2015, PiattiAPL2017}. In the former case, the gate-induced electric field breaks inversion symmetry and induces a finite Zeeman-like spin-orbit splitting in the conduction band \cite{KormanyosPRB2013, YuanPRL2014}, which in turn leads to spin-valley locking of the Cooper pairs and makes the gate-induced SC state extremely robust against in-plane magnetic fields (Ising SC) \cite{LuScience2015, SaitoNatPhys2016}. This intriguing gate-induced SC state was shown to survive down to the single-layer limit \cite{CostanzoNatNano2016, FuQuantMater2017}. Gated MoS\ped{2} may also develop CDW order for electron doping levels beyond the SC dome \cite{RosnerPRB2014, ZhuangPRB2017, PiattiApsusc2018_1}. Despite this intense investigation, the exact mechanism responsible for the onset of the SC state in gated MoS\ped{2} is still uncertain. Some theoretical works suggest that the electron-phonon coupling in doped MoS\ped{2} can become large enough to support a conventional phonon-driven SC pairing \cite{RosnerPRB2014, GePRB2013, DasPRB2015}. However, Costanzo et al. recently reported hints of non-trivial SC pairing from tunnelling spectra \cite{CostanzoNatNano2018}, which may require more exotic pairing mechanism to be explained \cite{YuanPRL2014, RoldanPRB2013, KhezerlouPRB2016, NakamuraPRB2017, HsuNatCommun2017}. This picture may be further complicated by the reconstruction of the conduction band below the Fermi level due to the formation of polarons \cite{KangNatMater2018}.

A common feature among the predicted SC phases in electron-doped $2H$-MoS\ped{2} is that their features hinge on the geometry of the Fermi surface, which in general can be composed by two electron pockets at K/K\apex{\prime}, six electron pockets at Q/Q\apex{\prime}, or both (see Fig.\ref{figure:material_structure}b). Which conduction band minima are pupulated, and whether the bands are spin-orbit split, depends on the number of layers, field-effect doping, and strain \cite{BrummePRB2015, BrummePRB2016}. Experimentally, this can lead to conflicting results depending on sample preparation and probing technique \cite{CuiNatNano2015, WuNatCommun2016, ChenPRL2017, KangNanoLett2017, PiattiNanoLett2018, PisoniNanoLett2017}. In Ref.\onlinecite{PiattiNanoLett2018}, we measured the transconductance of FETs realized on ion-gated MoS\ped{2} nanolayers to probe their Fermi surface \textit{in situ} by detecting the characteristic kinks associated to the onset of doping in high-energy sub-bands, an approach that was already employed successfully in gated few-layer graphene \cite{YePNAS2011, GonnelliSciRep2015, Gonnelli2dMater2017, PiattiApsusc2017}. Despite a qualitative agreement with the behavior predicted in Ref.\onlinecite{BrummePRB2016} for a three-layer (3L), the crossing of the high-energy sub-bands at Q/Q\apex{\prime} was observed at starkly lower doping levels. Since tensile strain can strongly change the energy difference between the two conduction band minima (see Fig.\ref{figure:material_structure}c), this difference was tentatively ascribed to the mismatch between the amount of strain measured in the real MoS\ped{2} device and that employed in the theoretical calculations, as well as the different number of layers.

In this work, we tackle this issue quantitatively by performing \textit{ab initio} Density Functional Theory (DFT) calculations on nanolayers of $2H$-MoS\ped{2} in the FET architecture, setting the number of layers and the amount of strain to those determined in real devices. Under these conditions, we show that the Fermi level crosses the bottom of the high-energy sub-bands at Q/Q\apex{\prime} at doping levels that agree quantitatively with those where the transconductance kinks are observed experimentally. We extend a simple 2D model to explicitly calculate the doping-dependence of the conductivity from the \textit{ab initio} bandstructure and use it to reproduce the thermal broadening of the kinks and the sub-band occupation at any temperature $T>0$. Overall, we demonstrate that experimental measurements of the transconductance and \textit{ab initio} DFT calculations of the bandstructure can be combined to precisely investigate the fermiology of gated TMDs, as well as to determine key physical quantities such as the amount of strain and the spin-orbit splitting between sub-bands. Our result will provide a useful guidance for the understanding of the intriguing SC properties of gated MoS\ped{2} and other layered materials.

\section{Computational details}

\begin{figure}[t]
\begin{center}
\includegraphics[keepaspectratio, width=0.9\columnwidth]{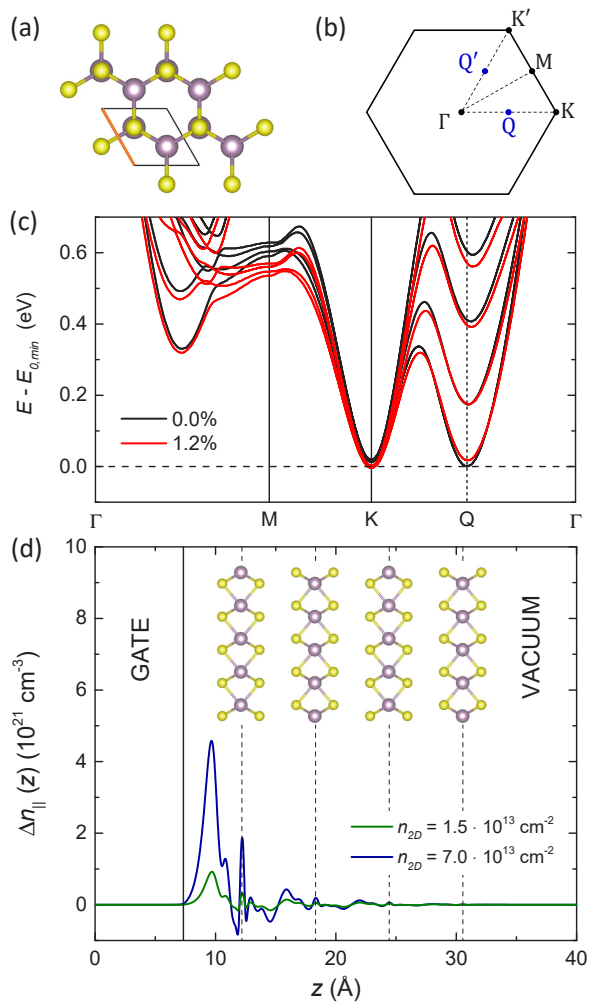}
\end{center}
\caption {
(a) Top view of the structure of MoS\ped{2} in the $2H$ crystal structure (greyish purple - Mo, yellow - S) \cite{VESTA}. The unit cell is indicated by the solid black line, the lattice parameter $a$ by the orange segment. 
(b) First Brillouin Zone of $2H$-MoS\ped{2}. High symmetry points $\Gamma$, K/K\apex{\prime} and M, as well as the Q/Q\apex{\prime} points, are highlighted. 
(c) Bandstructure of undoped 4L-MoS\ped{2} calculated with the bulk lattice parameter ($0.0$\% tensile strain, solid black line) or with the fully relaxed structure ($1.2$\% tensile strain, solid red line). The global minimum of the conduction band $E_{0,min}$ sits at Q/Q\apex{\prime} in the former case, at K/K\apex{\prime} in the latter.
{\color{blue}(d) Planar-averaged difference of the charge density in doped and undoped 4L-MoS\ped{2}, $\Delta n_{||}$, along the out-of-plane direction $z$, for two different values of $n_{2D}$. The vertical solid line indicates the position of the gate potential barrier. The vertical dashed lines mark the positions of the Mo atomic planes. A side view of the structure of 4L-MoS\ped{2} is also shown for clarity \cite{VESTA}.}
} \label{figure:material_structure}
\end{figure}

We computed the electronic bandstructure of \mbox{4L-MoS\ped{2}} by \textit{ab initio} DFT calculations as implemented in the Quantum ESPRESSO package \cite{GiannozziJPCM2017}. This number of layers was selected to match those of the samples employed in Ref.\onlinecite{PiattiNanoLett2018}. The valence-electron wave functions were expanded in the plane-wave basis set. Spin-orbit interaction was included via non-collinear calculations and through the use of full-relativistic projected augmented plane-wave (PAW) pseudopotentials \cite{BlochlPRB1994} for both atomic species. Correlations were included using the Perdew-Burke-Ernzerhof (PBE) exchange-correlation functional \cite{PerdewPRL1996} also taking into account the van-der-Waals dispersive corrections \cite{GrimmeJCC2006}. The cut-off energy for the plane-wave expansion and that for the density were set to $50$ Ry and $410$ Ry respectively, while the Brillouin zone integration was performed on a $64\times64\times1$ Monkhorst-Pack grid \cite{MonkhorstPRB1976} with a Gaussian smearing of $0.002$ Ry. The solution of the Kohn-Sham equations was obtained by setting the self-consistency condition on the total energy to $10^{-9}$ Ry, while the structure relaxation was checked upon convergence of the total force acting on the atoms set to $10^{-4}$ Ry/Bohr. 

The presence of the electric field in the FET configuration was taken into account through a recently implemented routine \cite{SohierPRB2017} where the metallic gate is modeled with a uniform planar distribution of charges of areal density $n_{2D}$, and the nanolayer is charged with an equal and opposite charge density (the field-induced doping charge) which maintains charge neutrality. In the following, we will consider only field-induced electron accumulation in the nanolayer ($n_{2D}>0$). The dielectric is substituted with a potential barrier of height equal to $2.0$ Ry that prevents unphysical charge spilling. {\color{blue}This routine allows to self-consistently determine the structural and electronic response of the gated system to the applied electric field from first principles, including the electrostatic screening arising both from the pristine electrons in the valence band and the field-induced electrons in the conduction band \cite{SohierPRB2017, BrummePRB2014} (see Fig.\ref{figure:material_structure}d). This approach has been employed to investigate the properties of a wide variety of layered materials upon field-effect doping, including graphene \cite{SohierPRB2017}, ZrNCl \cite{BrummePRB2014}, phosphorene and arsenene \cite{SohierArXiv2018}, and various semiconducting TMDs with thickness between 1 and 3 layers \cite{BrummePRB2015, BrummePRB2016, SohierArXiv2018}. Moreover, we explicitly employed it to assess the anomalous electrostatic screening of gated NbN thin films in presence of high electric fields \cite{PiattiApsusc2018_2}, obtaining an excellent agreement with the experimental results \cite{PiattiPRB2017}.}

The nanolayer was first relaxed at $n_{2D}=0$ to find the equilibrium value $a_{rel}$ of the in-plane lattice parameter $a$. Starting from the experimentally-measured bulk value, $a_{exp}=3.160\,\mathrm{\AA}$ \cite{StrainBulk}, we obtained \mbox{$a_{rel} = 3.198$ \AA}, consistent with Refs.\onlinecite{BrummePRB2015, BrummePRB2016}. For each value of $n_{2D}\neq0$, we then looked for the equilibrium position of the nanolayer with respect to the potential barrier due to the presence of the electric field. This second relaxation was carried out with $a=a_{rel}$ to allow for the correct minimization of the force acting on the atoms. Once the equilibrium distance between the barrier and the nanolayer for a given value of $n_{2D}$ was known, different amounts of tensile strain with respect to $a_{exp}$ were introduced by setting the appropriate value for $a$ \cite{BrummePRB2016}. This procedure was repeated for each value of $n_{2D}$.

{\color{blue}Finally, we note that in this work we do not attempt an \textit{ab initio} computation of the SC properties of gated MoS\ped{2}: in the following, when the doping dependence of the electronic bandstructure is discussed in relation to the doping dependence of $T_c$, the latter is obtained from the existing literature \cite{YeScience2012, LuScience2015, SaitoNatPhys2016, CostanzoNatNano2016, FuQuantMater2017, CostanzoNatNano2018, ChenPRL2017, PiattiNanoLett2018}. Assuming that SC in gated MoS\ped{2} is dominated by electron-phonon coupling -- a point which is currently debated \cite{DasPRB2015, RoldanPRB2013} -- a first-principles determination of its SC properties would at least entail the computation of the Fermi-surface average of either the electron-phonon matrix elements \cite{GePRB2013} or the electron-phonon spectral function \cite{UmmarinoPRB2017}. These in turn require the calculation of the phononic bandstructure as a function of field-effect doping within the framework of Density Functional Perturbation Theory \cite{SohierPRB2017, SohierArXiv2018}. This treatment of the system is significantly more complex than the one presented here and is left to future work. The direct investigation of the SC properties of gated MoS\ped{2} is therefore beyond the scope of our current approach, such as the calculation of $T_c$, of the SC proximity effect \cite{PiattiPRB2017, UmmarinoPRB2017, UmmarinoJPCM2019, SunAPL2014, PerconteNatPhys2018} which can become relevant in the dual-gate configuration \cite{ChenPRL2017}, of the exact spatial confinement of the superfluid, and of the symmetry of the SC order parameter.}

\section{Results}

In Fig.\ref{figure:material_structure}c we show the bandstructure of 4L-MoS\ped{2} computed at $n_{2D} = 0$ with no tensile strain ($a=a_{exp}$, solid black line) and $1.20$\% tensile strain, corresponding to the relaxed structure ($a=a_{rel}$, solid red line). It can clearly be seen that, in absence of a transverse electric field, the presence of a finite tensile strain is sufficient to shift the position of the global conduction band minimum between the K/K\apex{\prime} and Q/Q\apex{\prime} points. Specifically, with no tensile strain the global minimum sits at Q/Q\apex{\prime}, as is observed in the bulk. In the fully relaxed structure, the global minimum sits at K/K\apex{\prime} instead, as in the single-layer. Tensile strain can thus have a profound impact on the details of the bandstructure and must be selected carefully in order to match the experimental results.

\begin{figure}
\begin{center}
\includegraphics[keepaspectratio, height=0.89\textheight]{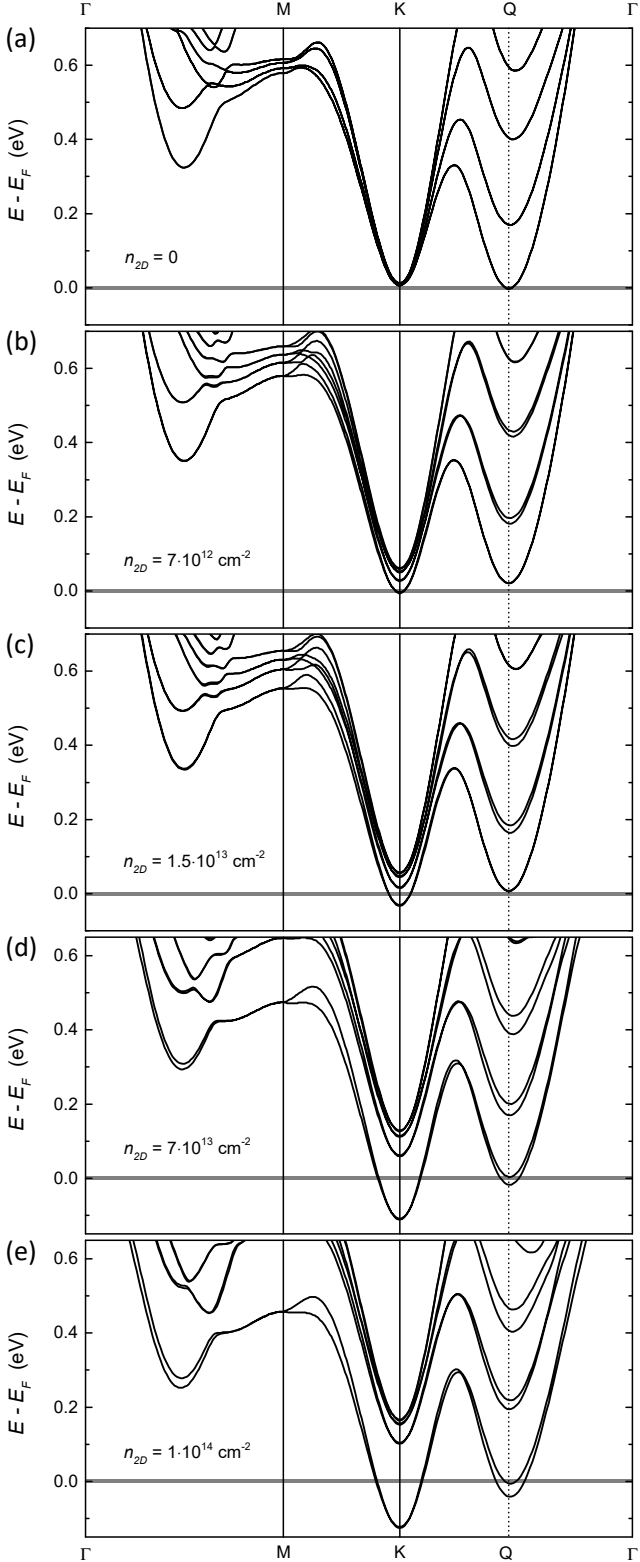}
\end{center}
\caption {Electronic bandstructure of 4L-MoS\ped{2} for increasing field-effect doping at $0.13$\% tensile strain. The position of the Fermi level is highlighted with an uncertainty of $10$ meV.
} \label{figure:bandstructure}
\end{figure}

We thus determined the amount of tensile strain necessary to reproduce the experimental observations reported in Ref.\onlinecite{PiattiNanoLett2018} in 4L-MoS\ped{2}. This was done by requiring the Fermi level to cross the bottom of the Q\ped{1} and Q\ped{2} sub-bands for values of $n_{2D}$ comparable to those where the conductivity kinks were observed experimentally ($n_{2D}|_{Q_1}\sim 1.5\cdot 10^{13}$ cm\apex{-2} and $n_{2D}|_{Q_2}\sim7\cdot 10^{13}$ cm\apex{-2}). Out of all the guess values we tested (see Supplementary Material \cite{Supplementary} for more details), a tensile strain of $0.13$\% ($a=3.164$ \AA) was found to best comply with the requirement, and has to be compared with the experimental value. In Ref.\onlinecite{PiattiNanoLett2018}, the total tensile strain in the MoS\ped{2} device was estimated to be $\sim 0.23$\% and due to two different contributions: the first one $\sim 0.13$\% was \textit{directly measured} by low-$T$ Raman spectroscopy, and is due to the thermal-expansion coefficient (TEC) mismatch between the MoS\ped{2} flake and the Au leads \cite{PiattiNanoLett2018}; the second contribution $\sim0.10$\% was instead \textit{estimated} from the TEC mismatch between the MoS\ped{2} flake and the SiO\ped{2} substrate, but not measured directly \cite{PiattiNanoLett2018}. As such, the excellent agreement of the \textit{directly-measured} tensile strain with the one determined by DFT suggests that no significant substrate-induced strain was present in the devices, likely owing to the weak van-der-Waals bond between the substrate and the MoS\ped{2} flakes.

{\color{blue}We directly assess the redistribution of the charge density in field-effect doped 4L-MoS\ped{2} due to the screening of the gate electric field as follows: First we calculate the planar-averaged charge density, $n_{||}$, with no field-induced charge [$n_{||}(n_{2D}=0)$] and with finite field-induced charge \mbox{[$n_{||}(n_{2D}\neq0)$]}. Both have to be computed with the relaxed structure of the doped system \cite{BrummePRB2014, PiattiApsusc2018_2}. We then calculate their difference $\Delta n_{||}$ for two values of $n_{2D}$ and plot it in Fig.\ref{figure:material_structure}d, namely $n_{2D}=n_{2D}|_{Q_1}$ (solid green line) and $n_{2D}=n_{2D}|_{Q_2}$ (solid blue line). $\Delta n_{||}$ represents the total screening charge -- which includes contributions both from the free carriers in the conduction band and the rearranged valence electrons \cite{BrummePRB2014, PiattiApsusc2018_2} -- and is positive where electrons are accumulated and negative where they are depleted. Consistently with the cases of 1L, 2L and 3L-MoS\ped{2} \cite{BrummePRB2015}, most of the screening charge is confined within the first layer, as expected for a strong population of the K/K\apex{\prime} valleys; the smaller amount of screening charge which can be observed in the second layer can be ascribed to the population of the Q/Q\apex{\prime} valleys instead \cite{BrummePRB2015}. For both values of $n_{2D}$, the screening charge is strongly asymmetric along the out-of-plane direction, and is thus responsible for the lifting of the inversion symmetry that would otherwise be present in any even-layered MoS\ped{2} nanolayer.}

In Fig.\ref{figure:bandstructure} we show the evolution of the bandstructure of \mbox{4L-MoS\ped{2}}, computed at the optimized tensile strain of $0.13$\%, with increasing electron doping. At $n_{2D}=0$ (Fig.\ref{figure:bandstructure}a), the global minimum of the conduction band falls at the Q/Q\apex{\prime} points, with the secondary minima at the K/K\apex{\prime} points lying only $6$ meV above the Fermi level. Given that the energy accuracy of DFT calculations is typically estimated to be $\sim 10$ meV \cite{DFTaccuracy}, the two minima can be considered to be nearly degenerate. With no applied electric field to break the inversion symmetry, both valleys are spin-degenerate. At $n_{2D} = 7\cdot10^{12}$ cm\apex{-2} (Fig.\ref{figure:bandstructure}b), the presence of a finite electric field shifts the global conduction band minimum to the K/K\apex{\prime} points, filling the corresponding electron pockets. The secondary minimum at Q/Q\apex{\prime} is lifted above the Fermi level and is not populated. At $n_{2D} = 1.5\cdot10^{13}$ cm\apex{-2} (Fig.\ref{figure:bandstructure}c), the K/K\apex{\prime} valleys are strongly populated and the Fermi level crosses the bottom of the Q\ped{1} sub-band in the Q/Q\apex{\prime} valleys. In both cases, the broken inversion symmetry leads to a small but finite spin-orbit splitting in both valleys (not seen in this scale; see later for details). At $n_{2D} = 7\cdot10^{13}$ cm\apex{-2} (Fig.\ref{figure:bandstructure}d), the Fermi level crosses the bottom of the Q\ped{2} sub-band, and at $n_{2D} = 1\cdot10^{14}$ cm\apex{-2} (Fig.\ref{figure:bandstructure}e) all the sub-bands are strongly populated; in both cases, the spin-orbit splitting increases significantly, especially at Q/Q\apex{\prime}. 

\begin{figure}
\begin{center}
\includegraphics[keepaspectratio, width=0.9\columnwidth]{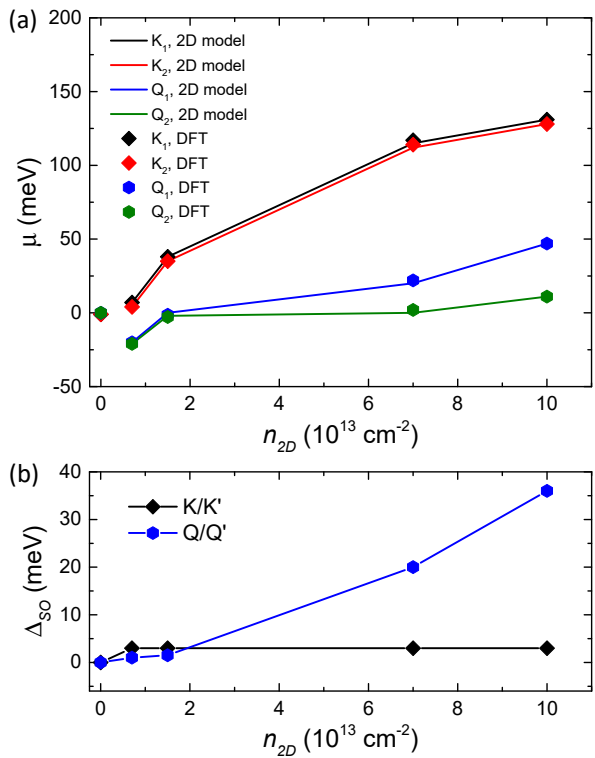}
\end{center}
\caption {(a) Chemical potentials in the four sub-bands for increasing field-effect doping. Symbols are the values directly obtained from the \textit{ab initio} calculations (black/red diamonds - K\ped{1}/K\ped{2}, blue/green hexagons - Q\ped{1}/Q\ped{2}). Solid lines are the linearized trends of the chemical potentials which are the input for the simplified 2D model (same color coding). (b) Spin-orbit splitting in the two valleys for increasing field-effect doping. Black diamonds refer to the splitting between K\ped{1} and K\ped{2}. Blue hexagons to that between Q\ped{1} and Q\ped{2}. Solid lines are guides to the eye.
} \label{figure:Fermi_and_SOS}
\end{figure}

We can gain a better quantitative insight in the doping-dependence of the bandstructure by calculating the chemical potential of each sub-band $\mu_i = E_F - E_{0,i}$ -- where $E_{0,i}$ is the bottom of the $i$-th sub-band -- and plotting it as a function of $n_{2D}$ (Fig.\ref{figure:Fermi_and_SOS}a). At $n_{2D}=0$ the Fermi level can be considered to lie just below the bottom of the conduction band minimum and $\mu_i \simeq0$ in all the sub-bands. For small values of field-effect doping, $\mu_\mathrm{K_1,K_2}$ strongly increases while $\mu_\mathrm{Q_1,Q_2}$ becomes negative, as the corresponding sub-bands are pushed above $E_F$: free electrons populate the K/K\apex{\prime} valleys only. When $E_F$ crosses Q\ped{1}, the rate at which $\mu_\mathrm{K_1,K_2}$ increases is reduced, since a significant amount of the field-induced electrons are absorbed by the Q/Q\apex{\prime} valleys; in this range of $n_{2D}$, the Q\ped{2} sub-band remains pinned below $E_F$, indicating that the Q/Q\apex{\prime} valleys are fully spin-polarized. When $E_F$ crosses Q\ped{2}, the rate at which $\mu_\mathrm{K_1,K_2}$ increases is further reduced, while that at which $\mu_\mathrm{Q_1,Q_2}$ increases is enhanced. At even larger values of $n_{2D}$, it is likely that $\mu_\mathrm{K_1,K_2}$ would start to decrease and eventually become negative, as predicted in 1L and 3L-MoS\ped{2} \cite{BrummePRB2015}.

From the knowledge of the chemical potentials, we can also directly obtain the doping dependence of the spin-orbit splitting in each valley as $\Delta_{SO}|_\mathrm{K/K^\prime} = \mu_\mathrm{K_1} - \mu_\mathrm{K_2}$ and $\Delta_{SO}|_\mathrm{Q/Q^\prime} = \mu_\mathrm{Q_1} - \mu_\mathrm{Q_2}$, which we plot in Fig.\ref{figure:Fermi_and_SOS}b. At $n_{2D} = 0$, no spin-orbit splitting is present in either valley, as expected due to the centrosymmetric structure of even-layered 2\textit{H}-MoS\ped{2}. At finite $n_{2D}$, the transverse electric field breaks the inversion symmetry and lifts the spin degeneracy, resulting in a finite spin-orbit splitting. Moreover, $\Delta_{SO}$ exhibits a starkly different behavior in the two valleys: in the K/K\apex{\prime} valleys, it almost immediately saturates to $\simeq 3$ meV and does not show a further doping dependence; in the Q/Q\apex{\prime} valleys, on the other hand, it increases in the whole doping range and eventually becomes much larger ($\simeq 36$ meV at $n_{2D}=1\cdot 10^{14}$ cm\apex{-2}). This behavior of $\Delta_{SO}$ is expected, since the degree to which inversion symmetry is broken in MoS\ped{2} increases with increasing transverse electric field \cite{LuScience2015, SaitoNatPhys2016, ChenPRL2017}. This in turn is associated to the fact that the gate-induced charge carriers tend to become more localized within the first layer from the surface as the electric field increases \cite{BrummePRB2015,RoldanPRB2013}, resulting in the gated multilayer to mimic the behavior of a doped monolayer on an insulating bulk \cite{LuScience2015, SaitoNatPhys2016, ChenPRL2017}.

\begin{figure}
\begin{center}
\includegraphics[keepaspectratio, width=0.95\columnwidth]{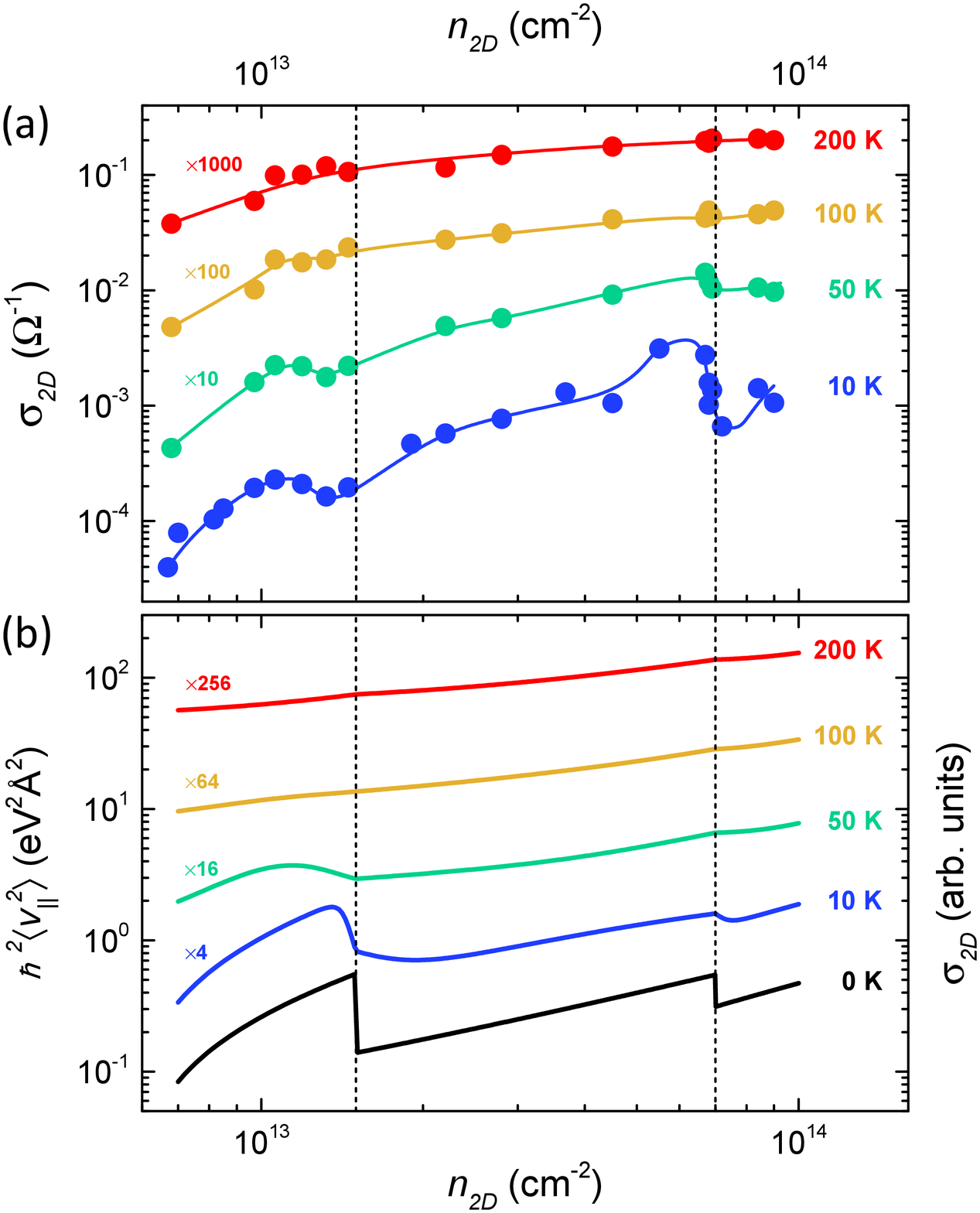}
\end{center}
\caption {(a) Sheet conductance of 4L-MoS\ped{2} for increasing field-effect doping, measured at different temperatures, in log-log scale. Filled circles are data adapted from Ref.\onlinecite{PiattiNanoLett2018}. Solid lines are guides to the eye. (b) Averaged squared velocity of 4L-MoS\ped{2} for increasing field effect doping, calculated with the 2D model at different temperatures. In both panels, the curves are shifted for clarity by multiplying each curve for a different constant. Vertical dashed lines highlight the doping values where the Fermi level crosses the bottom of Q\ped{1} and Q\ped{2} in the 2D model.
} \label{figure:kinks}
\end{figure}

We now show explicitly that the doping dependence of the bandstructure at $0.13$\% tensile strain is able to reproduce the experimentally observed kinks in the conductivity (shown in Fig.\ref{figure:kinks}a). This can be done by estimating the average squared in-plane velocity of the charge carriers, $£\langle v_\parallel^2 \rangle$, from the calculated bandstructure, since the sheet conductance $\sigma_{2D}$ is -- at a first approximation -- linearly proportional to $\langle v_\parallel^2 \rangle$ \cite{BrummePRB2016}. To estimate $\langle v_\parallel^2 \rangle$, we employ the simplified 2D model developed in Ref.\onlinecite{BrummePRB2016} at $T=0$, and extend it to finite $T$. This model approximates the exact momentum dependence of the energy sub-bands with an isotropic and parabolic dispersion, $\epsilon(k)=\hbar^2k^2/(2m_i)$, assuming a 2D density of states $g_i^vm_i/(\pi\hbar^2)$, where $g_i^v$ are the valley degeneracies and $m_i$ are the effective masses for each sub-band $i$. At $T=0$, the model has already been shown to reliably match the results from the full \textit{ab initio} calculations for the doping dependence of the chemical potentials, total density of states, and $\langle v_\parallel^2 \rangle$, at any doping level such that the non-parabolicity of the energy bands is negligible ($n_{2D}\lesssim 2\cdot10^{14}$ cm\apex{-2}) \cite{BrummePRB2016}. In the following, we will conservatively employ the model only up to $n_{2D}= 1\cdot10^{14}$ cm\apex{-2}.
Under the assumption outlined above, we can rewrite Eq.5 of Ref.\onlinecite{BrummePRB2016} as:
\begin{equation}
\langle v_\parallel^2 \rangle = \frac{\sum_i g_i^v \int_0^{\infty}  \left(-\frac{\partial f\left(\epsilon,\mu_i,T\right)}{\partial \epsilon}\right) \epsilon \mathrm{d}\epsilon  }{\sum_i g_i^v m_i \int_0^{\infty}  \left(-\frac{\partial f\left(\epsilon,\mu_i,T\right)}{\partial \epsilon}\right) \mathrm{d}\epsilon  }
\label{eq:velocity}
\end{equation}
where the sums run over each sub-band $i$ with chemical potential $\mu_i$ and Fermi-Dirac distribution function $f\left(\epsilon,\mu_i,T\right) = \left(\mathrm{exp}\left(\left(\epsilon-\mu_i\right)/\left(k_BT\right)\right)+1\right)^{-1}$ with $k_B$ being the Boltzmann constant. Since we know the dependence of $\mu_i$ on $n_{2D}$ from the \textit{ab initio} calculations, Eq.\ref{eq:velocity} allows us to determine $\langle v_\parallel^2 \rangle$ for any value of $T$ and $n_{2D}$. We benchmarked the accuracy of the 2D model in determining the effect of a finite $T$ against the values computed fully \textit{ab initio} with the BoltzTrap package \cite{BoltzTrap} in 1L and 3L-MoS\ped{2} \cite{BrummePRB2016}, finding excellent agreement (see Supplementary Material \cite{Supplementary} for details).

In Fig.\ref{figure:kinks}b, we plot $\hbar^2\langle v_\parallel^2 \rangle$, calculated from the values of $\mu_i$ reported in Fig.\ref{figure:Fermi_and_SOS}a, as a function of $n_{2D}$ for different values of $T$. As expected, the crossing of the high-energy sub-bands results in sharp kinks in $\langle v_\parallel^2 \rangle$ at $T=0$, which get progressively smeared out by increasing $T$. In particular, clear signatures of band crossing disappear above $\sim 100$ K. These results are to be compared with the values of $\sigma_{2D}$ as a function of $n_{2D}$ reported in Ref.\onlinecite{PiattiNanoLett2018}, which we plot in Fig.\ref{figure:kinks}a for the same values of $T$. It is clear that the thermal broadening at different $T$ is fully captured by the 2D model, while the height of each kink is not reproduced correctly (the kinks associated to the crossing of Q\ped{1} and Q\ped{2} are respectively overestimated and underestimated). The thermal broadening can be assessed quantitatively by comparing the width of each kink, $\Delta n$, in the model and the experiment. $\Delta n$ is defined as the difference in the values of $n_{2D}$ where the local maximum and minimum in the transconductance/average squared velocity are observed close to each sub-band crossing at a given $T$. For the crossing of Q\ped{1} at $10$ K, we obtain $\Delta n_{exp} \simeq 3.5\cdot 10^{12}$ cm\apex{-2} from the experimental transconductance and $\Delta n_{th} \simeq 4.5\cdot 10^{12}$ cm\apex{-2} from the averaged squared velocity; for the crossing of Q\ped{2}, also at $10$ K, we obtain $\Delta n_{exp} \simeq 0.9\cdot 10^{13}$ cm\apex{-2} and $\Delta n_{th} \simeq 0.7\cdot 10^{13}$ cm\apex{-2}. The broadening of the kinks in the experimental measurements is thus fully accounted for by the effect of a finite $T$, and further sources of smearing (such as doping inhomogeneity or quantum fluctuations) appear to be negligible. 

We now discuss the possible reasons for the mismatch in the height of the kinks between the model and the experimental data. The overestimation of the height of the kink at Q\ped{1} likely originates from the assumption, employed in the 2D model, of a sharp change in the slope of the chemical potentials at the sub-band crossing instead of a smooth joint \cite{BrummePRB2016, Supplementary}. Conversely, the underestimation of the height of the kink at Q\ped{2} cannot be ascribed only to a simple meshing issue, and points to a more general violation of the direct proportionality between $\sigma_{2D}$ and $\langle v_\parallel^2 \rangle$ -- which holds only when the scattering rate is directly proportional to the density of states \cite{BrummePRB2016}. The underestimation of the height of the kink at Q\ped{2} thus suggests that a significant enhancement in the scattering rate may occur at the crossing of Q\ped{2}, possibly due to the opening of interband and/or intervalley scattering channels \cite{YePNAS2011, GonnelliSciRep2015, PiattiApsusc2017, Gonnelli2dMater2017, PiattiNanoLett2018}.

\section{Discussion}

The good agreement between the calculated behavior of $\langle v_\parallel^2 \rangle$ and the experimentally observed dependence of $\sigma_{2D}$ as a function of $n_{2D}$ brings a solid theoretical support to the interpretation proposed in Ref.\onlinecite{PiattiNanoLett2018}. Our results highlight that the amount of tensile strain in TMD nanolayers has to be accounted for in a precise way to obtain a quantitative agreement between bandstructure calculations and experimental conditions. 

In the specific case of electron-doped 4L-MoS\ped{2}, the crossing of the two spin-orbit split sub-bands at Q/Q\apex{\prime} clearly separates its electronic structure into four different regimes. At very low doping, our calculations indicate that electrons preferentially fill the valleys at Q/Q\apex{\prime}. Such a regime is consistent with the degeneracy of the Landau levels observed experimentally in Refs.\onlinecite{CuiNatNano2015, WuNatCommun2016, ChenPRL2017, PisoniNanoLett2017}. The increase of the transverse electric field at slightly larger doping levels $n_{2D} \sim 7\cdot 10^{12}$ cm\apex{-2} induces a first Lifshitz transition in the system, depopulating the Q/Q\apex{\prime} pockets and filling the K/K\apex{\prime} pockets instead. These are the smallest doping levels typically obtained with the ionic gating technique, indicating that indeed the ``default'' Fermi surface of ion-gated MoS\ped{2} nanolayers qualitatively resembles that of the doped single-layer, as was proposed in Refs.\onlinecite{LuScience2015, SaitoNatPhys2016, ChenPRL2017}. Therefore, for $n_{2D} \lesssim 1.5\cdot10^{13}$ cm\apex{-2}, $\mu_i\geq0$ only for $i=\mathrm{K_1,K_2}$, the Fermi surface consists of two spin-polarized electron pockets in each of the two K/K\apex{\prime} valleys. At $n_{2D}\simeq 1.5\cdot10^{13}$ cm\apex{-2}, the Fermi level crosses the bottom of the Q\ped{1} sub-band and the Fermi surface undergoes a second Lifshitz transition to the third regime, where one spin-polarized electron pocket appears in each of the six Q/Q\apex{\prime} valleys ($\mu_i\geq0$ for $i=\mathrm{K_1,K_2,Q_1}$). A third Lifshitz transition occurs at $n_{2D}\simeq 7\cdot10^{13}$ cm\apex{-2}, when the Fermi level crosses the bottom of the Q\ped{2} sub-band and a second spin-polarized electron pocket is added to each of the six Q/Q\apex{\prime} valleys. Thus, only in this fourth regime all valleys host electron pockets with both spin polarizations ($\mu_i\geq0$ for $i=\mathrm{K_1,K_2,Q_1,Q_2}$).

This precise determination of the doping-dependence of the Fermi surface of gated MoS\ped{2} is of key importance in unveiling the mechanism behind the onset of SC in this system. This stems from the fact that the electron-phonon coupling is strongly boosted by the increase of the number of phonon branches contributing to the coupling when the high-energy electron pockets emerge at Q/Q\apex{\prime} \cite{GePRB2013, PiattiNanoLett2018}, and these in turn are significantly modified by the field-induced spin-orbit splitting in the sub-bands \cite{BrummePRB2015}. In particular, the population of the second spin-polarized sub-band at Q/Q\apex{\prime}, Q\ped{2}, appears to be critical for the $T_c$ increase in the first half of the SC dome of field-effect doped MoS\ped{2}, a feature that could possibly be shared across several semiconducting TMDs \cite{PiattiNanoLett2018}. {\color{blue}Among intrinsically superconducting TMDs, on the other hand, a detailed investigation of possible strain-dependent Fermi surface reconstructions may provide valuable hints as to why SC is weakened when bulk flakes of one compound are thinned to the single-layer (such as in NbSe\ped{2} \cite{FrindtPRL1972, delaBarreraNatCommun2018}), while it is strengthened in other compounds (such as in TaS\ped{2} \cite{delaBarreraNatCommun2018, NavarroNatCommun2016}). Moreover,} the exact population of the various spin-polarized sub-bands also plays a pivotal role in the emergence of the topologically non-trivial states \cite{YuanPRL2014, RoldanPRB2013, KhezerlouPRB2016, NakamuraPRB2017, HsuNatCommun2017} which can be supported by the Ising SC associated to the presence of spin-valley locked Cooper pairs \cite{LuScience2015, SaitoNatPhys2016}.

In addition to giving key insight into the fermiology of gated TMD-based devices, our results show that the experimental detection of kinks in the doping-dependence of their in-plane transconductance at low $T$ can be combined with \textit{ab initio} DFT calculations to estimate two key physical parameters, the amount of tensile strain and the spin-orbit splitting in different valleys, which are otherwise difficult to directly access experimentally. Moreover, our analysis revealed hints to a significant enhancement of the interband (and/or intervalley) scattering rate at the crossing of Q\ped{2}. Thus, it would be worth investigating whether and how such a feature relates to the concomitant increase in $T_c$ and the SC coupling. This will require a more complex treatment of the system, combining more detailed calculations of the electronic bandstructure with an \textit{ab initio} determination of the doping-dependent electron-phonon coupling in the FET configuration \cite{SohierPRB2017, SohierArXiv2018}, and is left to future work.

\section{Conclusions}
In summary, we performed \textit{ab initio} DFT calculations on strained MoS\ped{2} nanolayers in the field-effect transistor configuration. By combining an analytic 2D model with the bandstructure calculated at increasing electron doping, we obtained the doping dependence of the average squared in-plane velocity of the charge carriers, which exhibits sharp kinks in correspondence to the onset of doping in the high-energy sub-bands at Q/Q\apex{\prime}. We explicitly showed that, when the amount of strain, the temperature and the number of layers are set to the values measured in real samples, position and broadening of these kinks reproduce with a good accuracy the doping-dependent features in the experimentally measured conductivity. Our results allow to employ the combination of transconductance measurements and \textit{ab initio} calculations to reliably determine the Fermi surface topology of field-effect doped nanolayers of transition metal dichalcogenides, as well as the values of strain and spin-orbit splitting present in real devices. Our findings provide a helpful and broadly applicable tool to better understand the electronic structure and investigate the origin of superconductivity in gated MoS\ped{2} and other transition metal dichalcogenides.

\bigskip

\section*{Acknowledgments}
We thank M. Calandra for fruitful scientific discussions. Computational resources were provided by hpc@polito (http://www.hpc.polito.it).

\end{document}